\documentclass[10pt,twocolumn,twoside]{IEEEtran}
\usepackage{amssymb}
\usepackage{mathbbol}
\usepackage{graphicx}
\usepackage{amsmath}
\usepackage{array}
\newcommand{\PreserveBackslash}[1]{\let\temp=\\#1\let\\=\temp}
\newcolumntype{C}[1]{>{\PreserveBackslash\centering}p{#1}}
\newcolumntype{R}[1]{>{\PreserveBackslash\raggedleft}p{#1}}
\newcolumntype{L}[1]{>{\PreserveBackslash\raggedright}p{#1}}
\usepackage[usenames]{color}
\usepackage{colortbl,booktabs}
\usepackage{tabularx}
\usepackage{multicol}
\usepackage{booktabs}
\usepackage[Symbol]{upgreek}
\usepackage{subfigure}
\usepackage{bm}
\usepackage{cite}
\usepackage{balance}
\usepackage{threeparttable}
\usepackage{amsmath}
\usepackage{dcolumn}
\usepackage{multirow}
\usepackage{booktabs}
\usepackage{amsfonts}
\newcolumntype{d}[1]{D{.}{.}{#1}}
\usepackage{algorithm} 
\usepackage{algorithmic} 
\usepackage{amsmath}

\usepackage{enumerate}
\usepackage{balance}

\begin{document}

\bibliographystyle{IEEEtran} 

\title{Hybrid Precoding-Based Millimeter-Wave Massive MIMO-NOMA with Simultaneous Wireless Information and Power Transfer}

%

\author{\IEEEauthorblockN{Linglong Dai, Bichai Wang, Mugen Peng, and Shanzhi Chen}
\thanks{Manuscript received March 12, 2018; revised July 4, 2018; accepted August 22, 2018. Date of publication XX, 2018; date of current version XX, 2018.}
\thanks{This work was supported by the State Major Science and Technology Special Project under 2017ZX03001025-006, the National Natural Science Foundation of China for Outstanding Young Scholars (Grant No. 61722109), the National Natural Science Foundation of China (Grant No. 61571270), and the Royal Academy of Engineering through the UK-China Industry Academia Partnership Programme Scheme (Grant No. UK-CIAPP$\backslash$49). \emph{(Corresponding author: Shanzhi Chen)}}
\thanks{Linglong Dai and Bichai Wang are with the Tsinghua National Laboratory for Information Science and Technology (TNList) as well as the Department of Electronic Engineering, Tsinghua University, Beijing 100084, China (E-mail: daill@tsinghua.edu.cn, wbc15@mails.tsinghua.edu.cn). Mugen Peng is with the Key Laboratory of Universal Wireless Communications (Ministry of Education), Beijing University of Posts and Telecommunications, Beijing 100876, China (E-mail: pmg@bupt.edu.cn). Shanzhi Chen is with the State Key Lab of Wireless Mobile Communication, China Academy of Telecommunication Technology, Beijing 100191, China (email: chensz@datanggroup.cn).} %

}

\maketitle

\begin{abstract}

Non-orthogonal multiple access (NOMA) has been recently considered in millimeter-wave (mmWave) massive MIMO systems to further enhance the spectrum efficiency. In addition, simultaneous wireless information and power transfer (SWIPT) is a promising solution to maximize the energy efficiency. In this paper, for the first time, we investigate the integration of SWIPT in mmWave massive MIMO-NOMA systems. As mmWave massive MIMO will likely use hybrid precoding (HP) to significantly reduce the number of required radio-frequency (RF) chains without an obvious performance loss, where the fully digital precoder is decomposed into a high-dimensional analog precoder and a low-dimensional digital precoder, we propose to apply SWIPT in HP-based MIMO-NOMA systems, where each user can extract both information and energy from the received RF signals by using a power splitting receiver. Specifically, the cluster-head selection (CHS) algorithm is proposed to select one user for each beam at first, and then the analog precoding is designed according to the selected cluster heads for all beams. After that, user grouping is performed based on the correlation of users' equivalent channels. Then, the digital precoding is designed by selecting users with the strongest equivalent channel gain in each beam. Finally, the achievable sum rate is maximized by jointly optimizing power allocation for mmWave massive MIMO-NOMA and power splitting factors for SWIPT, and an iterative optimization algorithm is developed to solve the non-convex problem. Simulation results show that the proposed HP-based MIMO-NOMA with SWIPT can achieve higher spectrum and energy efficiency compared with HP-based MIMO-OMA with SWIPT.

\end{abstract}


\section{Introduction}\label{S1}

\IEEEPARstart MILLIMETER-wave (mmWave) massive MIMO has been considered as one of the promising techniques for 5G wireless communications, since it can provide wider bandwidth and achieve higher spectrum efficiency~\cite{mMIMO1}~\cite{mMIMO2}. It is well known that in conventional MIMO systems, each antenna usually requires one dedicated radio-frequency (RF) chain to realize the fully digital signal processing~\cite{RFMIMO1}~\cite{RFMIMO2}. In this way, the use of a very large number of antennas in mmWave massive MIMO systems leads to an equally large number of RF chains, which will result in unaffordable hardware cost and energy consumption~\cite{RFMIMO3}. To address this issue, hybrid precoding (HP) has been proposed to significantly reduce the number of required RF chains in mmWave massive MIMO systems without an obvious performance loss~\cite{HPMIMO1}. The key idea of HP is to decompose the fully digital precoder into a high-dimensional analog precoder (realized by the analog circuit) to increase the antenna array gain and a low-dimensional digital precoder (realized by a small number of RF chains) to cancel interference~\cite{HPMIMO2,HPMIMO3,HPMIMO4,HPMIMO5}. Usually, two typical HP architectures are adopted~\cite{RFMIMO3}: 1) Fully-connected architecture, where each RF chain is connected to all antennas; 2) Sub-connected architecture, where each RF chain is connected to only a subset of antennas. For general comparison, the fully-connected architecture can achieve higher spectrum efficiency, while the sub-connected architecture is expected to achieve higher energy efficiency~\cite{RFMIMO3}.

To further increase the spectrum efficiency, non-orthogonal multiple access (NOMA) has been recently considered in mmWave massive MIMO systems~\cite{HPNOMA1,HPNOMA2,HPNOMA3,HPNOMA4}. It has been shown that NOMA can significantly improve the spectrum efficiency compared to the conventional orthogonal multiple access (OMA) schemes~\cite{NOMA1,NOMA2,NOMA3,NOMA4}. By using NOMA, more than one user can be supported in each beam with the aid of intra-beam superposition coding and successive interference cancellation (SIC)~\cite{NOMA1}~\cite{NOMA2}, which is essentially different from conventional mmWave massive MIMO using one beam to serve only one user at the same time-frequency resources. Particularly, NOMA was applied to beamspace MIMO for the first time in~\cite{HPNOMA1}, which can be regarded as a low-complexity realization of HP, and power allocation was optimized to maximize the achievable sum rate. In addition, NOMA was also utilized in fully-connected HP architecture in~\cite{HPNOMA2}, and digital precoding was designed by modifying the conventional block diagonalization (BD) precoding scheme. Furthermore, a more sophisticated digital precoding design, i.e., minorization-maximization (MM) based precoding, was proposed in~\cite{HPNOMA3} to maximize the achievable sum rate. Besides, power allocation was optimized in~\cite{HPNOMA4} to maximize the energy efficiency of mmWave massive MIMO-NOMA systems, and an iterative algorithm was proposed to obtain the optimal power allocation.

In addition to improving the spectrum efficiency, energy efficiency is also one of the key performance indicators (KPIs) for 5G, which is expected to be 100 times compared with that of current 4G wireless communications~\cite{ITU2083}. To this end, simultaneous wireless information and power transfer (SWIPT), which was initially proposed in~\cite{SWIPT1}, has attracted great interests in recent years~\cite{SWIPT2,SWIPT3,SWIPT4,SWIPT10,SWIPT11}. The key idea of SWIPT is that both information and energy could be extracted from the same received RF signals, which can be realized by power splitting receivers in practice~\cite{SWIPT5}. With the help of SWIPT, the battery-powered wireless communication devices can harvest energy from the RF signals to prolong their lifetime, which provides the potential to explore more energy-efficient networks, especially for Internet of Things (IoT) with millions of wireless devices~\cite{SWIPT6}. However, the trade-off between information rate and harvested energy level should be carefully considered to facilitate efficient SWIPT in multi-user systems, since inter-user interferences are usually harmful for the information decoding (ID), while they can be useful for energy harvesting (EH)~\cite{SWIPT2}.

In fact, some efforts have been endeavored to address this problem. Particularly, a joint transmit beamforming and power splitting optimization was investigated in~\cite{SWIPT7}, where the transmit power was minimized under the signal-to-interference-and-noise ratio (SINR) and EH quality of service (QoS) constraints for multi-user MIMO systems. In addition, the joint transceiver and power splitting design for downlink multi-user MIMO SWIPT networks was also investigated based on the mean square error (MSE) criterion in~\cite{SWIPT8}. For multi-cell multi-user downlink SWIPT systems, the joint transceiver and power splitting design was studied to optimize the energy efficiency in~\cite{SWIPT9}. Although SWIPT has the potential to realize energy efficient wireless communications, and has been investigated in some multi-user systems, the application of SWIPT in mmWave massive MIMO-NOMA systems has not been considered in the literature to the best of our knowledge, where new challenges will arise for the joint transceiver and power splitting optimization.

In this paper, we propose to integrate SWIPT in HP-based mmWave massive MIMO-NOMA systems to realize the spectrum- and energy-efficient wireless communications\footnote{Simulation codes are provided to reproduce the results presented in this paper: http://oa.ee.tsinghua.edu.cn/dailinglong/publications/publications.html.}. Particularly, a power splitting receiver is used for each user to achieve SWIPT by dividing the received signal into two parts for simultaneous information retrieval and energy storage. In such an system, we investigate the joint optimization of transceiver for ID and power splitting for EH, including user grouping, hybrid precoding design, power allocation, and power splitting factors design. Specifically, the contributions of this paper can be summarized as follows.

\begin{enumerate}

\item We propose to integrate SWIPT in HP-based mmWave massive MIMO-NOMA systems, including both the fully-connected architecture and sub-connected architecture. To the best of our knowledge, it is the first time to consider SWIPT in massive MIMO-NOMA systems. On the one hand, HP architecture at the base station (BS) can significantly reduce the number of required RF chains without an obvious performance loss, which can largely save the energy consumption at the BS, while guarantee the spectrum efficiency of mmWave massive MIMO systems. On the other hand, by using SWIPT, users can harvest energy from the received RF signals to prolong their life, which makes it more energy-efficient at the users. Note that although the application of SWIPT in conventional MIMO systems have been studied~\cite{SWIPT2,SWIPT3,SWIPT4,SWIPT7,SWIPT8}, the introduction of NOMA will result in additional challenges for the joint transceiver and power splitting optimization.

\item To enable the HP-based mmWave massive MIMO-NOMA systems with SWIPT, we investigate the joint transceiver and power splitting optimization. Specifically, the cluster-head selection (CHS) algorithm is proposed to select one user for each beam, and then the analog precoding is designed to obtain the antenna array gain according to the selected cluster heads for all beams. After that, user grouping is performed based on the equivalent channel correlation of the remaining users and the cluster-heads. Then, the digital precoding is designed to cancel the inter-user interference by selecting users with the strongest equivalent channel gain in each beam. Finally, the achievable sum rate is maximized by jointly optimizing power allocation as well as power splitting factors, which is very difficult to obtain the optimal solutions due to the coupling of different users' power allocation factors as well as the power splitting factors. To address this issue, an iterative optimization algorithm is developed to solve the non-convex problem, and the convergence and computational complexity are also analyzed.

\item The performance in terms of both spectrum efficiency and energy efficiency of the proposed HP-based mmWave massive MIMO-NOMA systems with SWIPT is evaluated by simulations. The convergence of the developed iterative optimization algorithm for joint power allocation and power splitting is validated, and it is shown that only 10 times of iteration are required to make it converged. Furthermore, we show that the proposed mmWave massive MIMO-NOMA systems with SWIPT can achieve higher spectrum and energy efficiency than those of mmWave massive MIMO-OMA systems with SWIPT.

\end{enumerate}

The rest of this paper is organized as follows. The system model of HP-based mmWave massive MIMO-NOMA system with SWIPT is introduced in Section II. User grouping and hybrid precoding design are discussed in Section III. In Section IV, the joint power allocation and power splitting optimization problem is formulated to maximize the achievable sum rate, under the achievable rate QoS and EH QoS constraints for each user. Furthermore, an iterative optimization algorithm is proposed to solve the non-convex problem. Simulation results are provided in Section V. Finally, conclusions are drawn in Section VI.

\emph{Notation}: We use upper-case and lower-case boldface letters to denote matrices and vectors, respectively; $(\cdot)^T$, $(\cdot)^H$, $(\cdot)^{-1}$, and $\|\cdot\|_p$ denote the transpose, conjugate transpose, matrix inversion, and $l_p$ norm operation, respectively. ${\rm E}\left\{  \cdot  \right\}$ denotes the expectation. $|\Gamma|$ denotes the number of elements in set $\Gamma$. ${\bf{A}}{\left( {i,:} \right)_{i \in \Gamma }}$ denotes the submatrix of $\bf{A}$ that consists of the $i$th row of $\bf{A}$ for all ${i \in \Gamma }$, while ${\bf{A}}{\left( {:,j} \right)_{j \in \Gamma }}$ denotes the submatrix of $\bf{A}$ that consists of the $j$th column of $\bf{A}$ for all ${j \in \Gamma }$. We use the notation ${\cal{CN}}\left( \bf{m}, \bf{R} \right)$ to denote the complex Gaussian distribution with mean $\bf{m}$ and covariance $\bf{R}$, and $\mathcal{U}\left( { a ,b } \right)$ to denote the uniform distribution in the range $(a,b)$. $\otimes$ denotes the kronecker product. Finally, $\mathbf{I}_N$ is the $N \times N$ identity matrix, and $\Phi $ denotes the empty set.


\section{System Model}\label{S2}

In this paper, we consider a single-cell downlink mmWave massive MIMO-NOMA system, where the BS is equipped with $N$ antennas and $N_{\rm RF}$ RF chains to simultaneously serve $K$ single-antenna users~\cite{HPNOMA2,HPNOMA3,HPNOMA4}, and each user is equipped with a power splitting receiver for SWIPT.

\begin{figure}[tp]
\begin{center}
\includegraphics[width=0.7\linewidth]{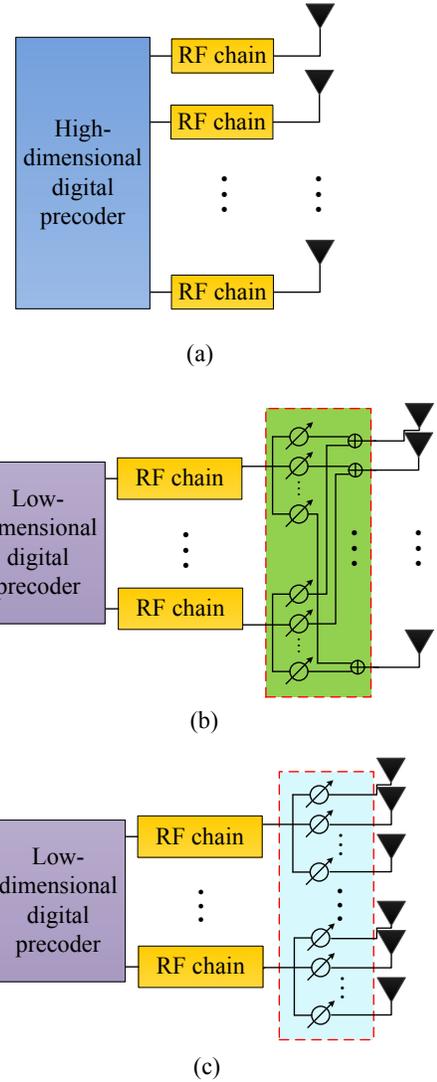} \caption{System models of mmWave MIMO architectures: (a) Fully digital MIMO; (b) Fully-connected HP architecture; (c) sub-connected HP architecture.}
\end{center}
\end{figure}

Fig. 1 shows three architectures of mmWave massive MIMO systems, i.e., the fully digital MIMO as shown in Fig. 1 (a), the fully-connected HP architecture as shown in Fig. 1 (b), and the sub-connected HP architecture as shown in Fig. 1 (c). From Fig. 1, we can see that for the fully digital MIMO, each antenna requires one dedicated RF chain, and thus the number of RF chains is equal to the number of antennas, which results in unaffordable energy consumption and hardware cost. On the contrary, the number of RF chains in HP architectures is less than the number of antennas, which can be realized by a high-dimensional analog precoder and a low-dimensional digital precoder. Specifically, for the fully-connected HP architecture in Fig. 1 (b), each of the $N_{\rm RF}$ RF chain is connected to all $N$ antennas by finite-resolution phase shifters, where $NN_{\rm RF}$ phase shifters are required, and thus the full array gain can be exploited by every RF chain. For the sub-connected HP architecture in Fig. 1 (c), each RF chain is connected to only a subset of $N$ BS antennas, so only $N$ phase shifters are required. In general, the sub-connected architecture is easier to be implemented and will likely be more energy efficient, while it may suffer some performance loss compared to the fully-connected architecture. In this paper, both of the fully-connected and sub-connected architectures will be considered.


In HP-based mmWave massive MIMO systems, the number of beams cannot exceed the number of RF chains, and each beam can only support one user at most~\cite{HPMIMO2}. Therefore, to fully achieve the multiplexing gain, we assume that the number of beams $G$ is equal to the number of RF chains $N_{\rm RF}$, i.e., $G = N_{\rm RF}$. On the contrary, each beam can support more than one user by using NOMA. Let $S_g$ for $g = 1,2, \cdots ,G$ denote the set of users served by the $g$th beam with $| {{S_g}} | \ge 1$, and we have ${S_i} \cap {S_j} = \Phi $ for $i \ne j$ as well as $\sum\limits_{g = 1}^{G} {| {{S_g}} |}  = K$. Then, the received signal at the $m$th user in the $g$th beam can be modeled as
\begin{equation}
\label{eq1}
\begin{array}{l}
\begin{aligned}
{y_{g,m}} =& {\bf{h}}_{g,m}^H{\bf{A}}\sum\limits_{i = 1}^G {\sum\limits_{j = 1}^{\left| {{S_i}} \right|} {{{\bf{d}}_i}\sqrt {{p_{i,j}}} {s_{i,j}}} }  + {v_{g,m}} \\
=& \underbrace {{\bf{h}}_{g,m}^H{\bf{A}}{{\bf{d}}_g}\sqrt {{p_{g,m}}} {s_{g,m}}}_{\rm desired \ signal} \\
&+ \underbrace {{{\bf{h}}_{g,m}^H{\bf{A}}{{\bf{d}}_g}\left( {\sum\limits_{j = 1}^{m - 1} {\sqrt {{p_{g,j}}} {s_{g,j}}}  + \sum\limits_{j = m + 1}^{\left| {{S_g}} \right|} {\sqrt {{p_{g,j}}} {s_{g,j}}} } \right)} }_{{\rm intra-beam \ interferences}} \\
&+ \underbrace {{\bf{h}}_{g,m}^H{\bf{A}}\sum\limits_{i \ne g} {\sum\limits_{j = 1}^{\left| {{S_i}} \right|} {{{\bf{d}}_i}\sqrt {{p_{i,j}}} {s_{i,j}}} } }_{{\rm inter-beam \ interferences}} + \underbrace {{v_{g,m}}}_{{\rm{noise}}},
\end{aligned}
\end{array}
\end{equation}
where ${s_{g,m}}$ is the transmitted signal with $E\{ {{{| {{s_{g,m}}} |}^2}} \} = 1$, ${{p_{g,m}}}$ is the transmitted power for the $m$th user in the $g$th beam, ${v_{g,m}}$ is the noise following the distribution ${\cal{CN}}\left( 0,{\sigma _v^2} \right)$, ${{{\bf{d}}_g}}$ is the $N_{\rm RF} \times 1$ digital precoding vector for the $g$th beam, ${\bf{A}}$ of size $N \times N_{\rm RF}$ is the analog precoding matrix, and we have ${\left\| {{\bf{A}}{{\bf{d}}_g}} \right\|_2} = 1$ for $g = 1,2, \cdots ,G$. Particularly, for the fully-connected architecture, the analog precoding matrix ${{\bf{A}}^{\left( {\rm{full}} \right)}}$ can be expressed as
\begin{equation}\label{eq2}
{{\bf{A}}^{\left( {\rm{full}} \right)}} = \left[ {{\bf{\bar a}}_1^{\left( {\rm{full}} \right)},{\bf{\bar a}}_2^{\left( {\rm{full}} \right)}, \cdots ,{\bf{\bar a}}_{{N_{\rm{RF}}}}^{\left( {\rm{full}} \right)}} \right],
\end{equation}
where the elements of ${{\bf{\bar a}}_n^{\left( {\rm{full}} \right)}} \in {^{N \times 1}}$ for $n = 1,2, \cdots ,{N_{\rm{RF}}}$ have the same amplitude ${1 / {\sqrt N }}$ but different phases~\cite{HPMIMO1}. For the sub-connected architecture, the analog precoding matrix ${{\bf{A}}^{\left( {\rm{sub}} \right)}}$ is
\begin{equation}\label{eq3}
{{\bf{A}}^{\left( {\rm{sub}} \right)}} = \left[ {\begin{array}{*{20}{c}}
{{\bf{\bar a}}_1^{\left( {\rm{sub}} \right)}}&{\bf{0}}& \cdots &{\bf{0}}\\
{\bf{0}}&{{\bf{\bar a}}_2^{\left( {\rm{sub}} \right)}}&{}&{\bf{0}}\\
 \vdots &{}& \ddots & \vdots \\
{\bf{0}}&{\bf{0}}& \cdots &{{\bf{\bar a}}_{{N_{\rm{RF}}}}^{\left( {\rm{sub}} \right)}}
\end{array}} \right].
\end{equation}
Without loss of generality, we assume that $M = {N \mathord{\left/ {\vphantom {N {{N_{RF}}}}} \right. \kern-\nulldelimiterspace} {{N_{\rm{RF}}}}}$ is an integer, and each RF chain is connected to $M$ antennas in the sub-connected architecture. Then, the elements of ${{\bf{\bar a}}_n^{\left( {\rm{sub}} \right)}} \in {^{M \times 1}}$ for $n = 1,2, \cdots ,{N_{\rm{RF}}}$ have the same amplitude ${1 / {\sqrt M }}$~\cite{HPMIMO2}~\cite{HPMIMO3}.

For the $N \times 1$ channel vector ${{\bf{h}}_{g,m}}$ of the $m$th user in the $g$th beam, we consider the widely used mmWave MIMO channel model as shown below~\cite{HPMIMO1,HPMIMO2,HPMIMO3}:
\begin{equation}\label{eq4}
{{\bf{h}}_{g,m}} = \sqrt {\frac{N}{{{L_{g,m}}}}} \sum\limits_{l = 1}^{{L_{g,m}}} {\alpha _{g,m}^{\left( l \right)}{\bf{a}}\left( {\varphi _{g,m}^{\left( l \right)},\theta _{g,m}^{\left( l \right)}} \right)},
\end{equation}
where ${{L_{g,m}}}$ denotes the number of paths for the $m$th user in the $g$th beam. ${\alpha _{g,m}^{\left( l \right)}}$ is the complex gain of the $l$th path. ${\varphi _{g,m}^{\left( l \right)}}$ and ${\theta _{g,m}^{\left( l \right)}}$ are the azimuth angle of departure (AoD) and elevation AoD of the $l$th path, and ${{\bf{a}}( {\varphi _{g,m}^{( l )},\theta _{g,m}^{( l )}} )}$ presents the $N \times 1$ array steering vector. Particularly, for the typical uniform linear array (ULA) with $N_1$ elements in horizon and $N_2$ elements in vertical, where $N = {N_1}{N_2}$~\cite{HPMIMO2}, we have
\begin{equation}\label{eq5}
{\bf{a}}\left( {\varphi ,\theta } \right) = {{\bf{a}}_{\rm{az}}}\left( \varphi  \right) \otimes {{\bf{a}}_{\rm{el}}}\left( \theta  \right),
\end{equation}
where ${{\bf{a}}_{\rm{az}}}\left( \varphi  \right) = \frac{1}{{\sqrt {{N_1}} }}{\left[ {{e^{j2\pi i\left( {{{{d_1}} \mathord{\left/ {\vphantom {{{d_1}} \lambda }} \right. \kern-\nulldelimiterspace} \lambda }} \right)\sin \left( \varphi  \right)}}} \right]_{i \in J\left( {{N_1}} \right)}}$, ${{\bf{a}}_{\rm{el}}}\left( \theta  \right) = \frac{1}{{\sqrt {{N_2}} }}{\left[ {{e^{j2\pi j\left( {{{{d_2}} \mathord{\left/ {\vphantom {{{d_2}} \lambda }} \right. \kern-\nulldelimiterspace} \lambda }} \right)\sin \left( \theta  \right)}}} \right]_{j \in J\left( {{N_2}} \right)}}$, $J\left( n \right) = \left\{ {0,1, \cdots ,n - 1} \right\}$, $\lambda $ is the signal wavelength, $d_1$ is the horizontal antenna spacing, and $d_2$ is the vertical antenna spacing. In mmWave communications, we usually have ${d_1} = {d_2} = {\lambda  \mathord{\left/ {\vphantom {\lambda  2}} \right. \kern-\nulldelimiterspace} 2}$~\cite{HPMIMO2}.

With the aid of power splitting receiver, the received signal at each user will be divided into two parts. One part is forwarded to the information decoder for ID, and the other is processed for EH~\cite{SWIPT8}. Let ${\beta _{g,m}}$, where $0 < {\beta _{g,m}} < 1$, denote the power splitting factor for the $m$th user in the $g$th beam, then the signal for EH can be represented as
\begin{equation}\label{eq6}
y_{g,m}^{\rm{EH}} = \sqrt {1 - {\beta _{g,m}}} {{y}_{g,m}},
\end{equation}
and the harvested energy is~\cite{SWIPT8}
\begin{equation}\label{eq7}
P_{g,m}^{\rm{EH}} = \eta \left( {1 - {\beta _{g,m}}} \right)\left( {\sum\limits_{i = 1}^G {\sum\limits_{j = 1}^{\left| {{S_i}} \right|} {\left\| {{\bf{\bar h}}_{g,m}^H{{\bf{d}}_i}} \right\|_2^2{p_{i,j}}} }  + \sigma _v^2} \right),
\end{equation}
where ${\bf{\bar h}}_{g,m}^H = {\bf{h}}_{g,m}^H{\bf{A}}$ is the equivalent channel vector, and $0 \le \eta  \le 1$ is the energy conversion efficiency. In the meanwhile, the signal for ID at the $m$th user in the $g$th beam can be expressed as
\begin{equation}\label{eq8}
y_{g,m}^{\rm{ID}} = \sqrt {{\beta _{g,m}}} {{y}_{g,m}} + {u_{g,m}},
\end{equation}
where ${u_{g,m}}$ is the noise caused by the power splitter following the distribution ${\cal{CN}}( 0,{\sigma _u^2} )$.

By using NOMA in each beam, intra-beam superposition coding at the transmitter and SIC at the receiver are performed. Without loss of generality, we assume that ${\| {{\bf{\bar h}}_{g,1}^H{{\bf{d}}_g}} \|_2} \ge {\| {{\bf{\bar h}}_{g,2}^H{{\bf{d}}_g}} \|_2} \ge  \cdots  \ge {\| {{\bf{\bar h}}_{g,| {{S_g}} |}^H{{\bf{d}}_g}} \|_2}$ for $g = 1,2, \cdots ,G$. Then, the $m$th user in the $g$th beam can remove the interference from the $j$th user (for all $j > m$) in the $g$th beam by performing SIC~\cite{NOMA1}, and the remaining received signal for ID at the $m$th user in the $g$th beam can be rewritten as
\begin{equation}
\label{eq9}
\begin{array}{l}
\begin{aligned}
\hat y_{g,m}^{\rm{ID}} =& \sqrt {{\beta _{g,m}}} \left( {{\bf{\bar h}}_{g,m}^H{{\bf{d}}_g}\sqrt {{p_{g,m}}} {s_{g,m}} + {\bf{\bar h}}_{g,m}^H{{\bf{d}}_g}\sum\limits_{j = 1}^{m - 1} {\sqrt {{p_{g,j}}} {s_{g,j}}}} \right. \\
& \left. {+ {\bf{\bar h}}_{g,m}^H\sum\limits_{i \ne g} {\sum\limits_{j = 1}^{\left| {{S_i}} \right|} {{{\bf{d}}_i}\sqrt {{p_{i,j}}} {s_{i,j}}} }  + {v_{g,m}}} \right) + {u_{g,m}}.
\end{aligned}
\end{array}
\end{equation}

Then, according to (\ref{eq9}), the SINR at the $m$th user in the $g$th beam can be written as
\begin{equation}\label{eq10}
{\gamma _{g,m}} = \frac{{\left\| {{\bf{\bar h}}_{g,m}^H{{\bf{d}}_g}} \right\|_2^2{p_{g,m}}}}{{{\xi _{g,m}}}},
\end{equation}
where
\begin{equation}\label{eq11}
\begin{array}{l}
\begin{aligned}
{\xi _{g,m}} =& \left\| {{\bf{\bar h}}_{g,m}^H{{\bf{d}}_g}} \right\|_2^2\sum\limits_{j = 1}^{m - 1} {{p_{g,j}}}  + \sum\limits_{i \ne g} {\left\| {{\bf{\bar h}}_{g,m}^H{{\bf{d}}_i}} \right\|_2^2} \sum\limits_{j = 1}^{\left| {{S_i}} \right|} {{p_{i,j}}} \\
& + \sigma _v^2 + \frac{{\sigma _u^2}}{{{\beta _{g,m}}}}.
\end{aligned}
\end{array}
\end{equation}

As a result, the achievable rate at the $m$th user in the $g$th beam is
\begin{equation}\label{eq12}
{R_{g,m}} = {\log _2}\left( {1 + {\gamma _{g,m}}} \right).
\end{equation}
Finally, the achievable sum rate is
\begin{equation}\label{eq13}
{R_{\rm sum}} = \sum\limits_{g = 1}^{G} {\sum\limits_{m =
1}^{\left| {{S_g}} \right|} {{R_{g,m}}} } ,
\end{equation}
which can be improved by carefully designing user grouping, analog precoding matrix ${\bf{A}}$, digital precoding $\left\{ {{{\bf{d}}_g}} \right\}_{g = 1}^{G}$, power allocation $\left\{ {{p_{g,m}}} \right\}_{g = 1,m = 1}^{G, \left| {{S_g}} \right|}$, and power splitting factors $\left\{ {{\beta _{g,m}}} \right\}_{g = 1,m = 1}^{G, \left| {{S_g}} \right|}$. Since it is very difficult to simultaneously obtain the optimal solutions for all these design parameters, we consider to design user grouping and hybrid precoding at first as described in the next section. Then, joint optimization of power allocation and power splitting will be introduced in Section IV.

\section{User Grouping and Hybrid Precoding}\label{S3}

We know that the number of users $K$ is larger than the number of RF chains $N_{\rm RF}$ in the considered system, while only $N_{\rm RF}$ different analog precoding vectors are available at the same time~\cite{HPMIMO3}. Therefore, to enable hybrid precoding, we propose the CHS algorithm to select one user for each beam (there are $G = N_{\rm{RF}}$ beams), and then the analog precoding is designed to obtain the antenna array gain according to the selected cluster heads for all beams. After that, user grouping is performed based on the equivalent channel correlation between the remaining users and the cluster-heads. Then, the digital precoding is designed to cancel inter-user interference by selecting users with the strongest equivalent channel gain in each beam.

\subsection{The proposed CHS algorithm}
To improve the system performance, we propose to select the cluster head for each beam by minimizing the channel correlation of the selected cluster heads. In this way, users in different beams will enjoy low channel correlation, which is beneficial for inter-beam interference cancellation.

In the proposed CHS algorithm, an adaptive threshold $\delta $ is introduced to measure the channel correlation of the cluster heads. Specifically, the user with the highest channel gain is selected as the cluster head for the first beam, and then users whose channel correlation with the first selected user is less than the threshold $\delta $ will be considered as the cluster head candidates for other beams. Particularly, the user with the highest channel gain out of the cluster head candidates is selected as the cluster head for the second beam. After that, the cluster head candidates will be updated by selecting users whose channel correlation with the second selected user is less than the threshold. This procedure is repeated until there is no candidate. Next, the threshold is updated by adding a small increment, and then the cluster head candidates are obtained by selecting users whose channel correlations with the previously selected cluster heads are less than the threshold. The threshold will be adaptively updated until the cluster heads are selected for all $G$ beams. The details of the proposed CHS algorithm are described in \textbf{Algorithm 1}, and the set of the selected cluster heads is denoted as $\Gamma $.

The proposed CHS algorithm enjoys the polynomial complexity. Specifically, in each iteration, the maximum complexity is $(2 + 2(K - 1))(K - 1)$ from step 9 to step 14, while the maximum complexity is $2(K - 1)$ from step 15 to step 18. Therefore, the complexity of Algorithm 1 is ${\cal O}(GK^2)$.

\begin{algorithm}[tp]
\caption{Proposed CHS algorithm}
\begin{algorithmic}[1]
\REQUIRE ~~\\
    The number of users $K$, and the number of beams $G$;
    Channel vectors: ${{\bf{h}}_{k}}$ for $k = 1,2, \cdots, K$;\\
    The initial threshold: $\delta $.
\ENSURE ~~\\
    The cluster head set $\Gamma $.

\STATE ${\bf{\Lambda }} = \left[ {{a_1},{a_2}, \cdots ,{a_K}} \right]$, where ${a_k} = {\left\| {{{\bf{h}}_k}} \right\|_2}$;

\STATE ${{{\bf{\tilde h}}}_k} = {{{{\bf{h}}_k}} \mathord{\left/ {\vphantom {{{{\bf{h}}_k}} {{a_k}}}} \right. \kern-\nulldelimiterspace} {{a_k}}}$ for $k = 1,2, \cdots, K$;

\STATE $\left[ {{\rm{\sim, }}{{\rm O}}} \right]{\rm{  =  sort}}\left( {{\bf{\Lambda }}{\rm{, 'descend'}}} \right)$;

\STATE $\Gamma  = {{\rm O}}\left( 1 \right)$;

\STATE ${\Gamma ^c} = {\rm O}/\Gamma $;

\STATE $\Omega  = {\Gamma ^c}$;

\STATE $g = 2$.

\WHILE {$g \le G$}

\IF {$\Omega  =  = \Phi $}

\WHILE {$\Omega  =  = \Phi $}

\STATE $\delta  = \delta  + {{\left( {1 - \delta } \right)} \mathord{\left/
 {\vphantom {{\left( {1 - \delta } \right)} {10}}} \right.
 \kern-\nulldelimiterspace} {10}}$;

\STATE $\Omega  = \left\{ {i \in {\Gamma ^c}\left| {\left| {{\bf{\tilde h}}_i^H{{{\bf{\tilde h}}}_j}} \right| < \delta ,\forall j \in \Gamma } \right.} \right\}$.

\ENDWHILE

\ENDIF

\STATE $\Omega  = \left\{ {i \in \Omega \left| {\left| {{\bf{\tilde h}}_i^H{{{\bf{\tilde h}}}_j}} \right| < \delta ,\forall j \in \Gamma } \right.} \right\}$;

\STATE $\Gamma  = \Gamma  \cup \Omega \left( 1 \right)$;

\STATE ${\Gamma ^c} = {\rm O}/\Gamma $;

\STATE $g=g+1$.

\ENDWHILE

\RETURN $\Gamma $.

\end{algorithmic}

\end{algorithm}

\subsection{Analog precoding}

In this paper, we consider the typical two-stage HP proposed in~\cite{HPMIMO4}\footnote{Note that more sophisticated HP schemes can be considered to further enhance the performance of mmWave massive MIMO-NOMA systems.}. The key idea of this scheme is to divide the HP design into two step, i.e., analog precoding and digital precoding. Particularly, for analog precoding, only quantized phase changes can be applied due to the practical constraints of phase shifters~\cite{HPMIMO5}. Considering $B$ bits quantized phase shifters, the non-zero elements of the fully-connected analog precoding matrix ${{\bf{A}}^{\left( {\rm{full}} \right)}}$ belong to
\begin{equation}\label{eq14}
\frac{1}{{\sqrt N }}\left\{ {{e^{j\frac{{2\pi n}}{{{2^B}}}}}:n = 0,1, \cdots, {2^B} - 1} \right\},
\end{equation}
while the non-zero elements of the sub-connected analog precoding matrix ${{\bf{A}}^{\left( {\rm{sub}} \right)}}$ belong to
\begin{equation}\label{eq15}
\frac{1}{{\sqrt M }}\left\{ {{e^{j\frac{{2\pi n}}{{{2^B}}}}}:n = 0,1, \cdots, {2^B} - 1} \right\}.
\end{equation}

Based on the cluster head set $\Gamma $ obtained in the previous subsection, the analog precoding can be designed according to the channel vectors of users in $\Gamma $. More particularly, the analog precoding vectors can be obtained by maximizing the array gains ${| {{\bf{h}}_{\Gamma ( g )}^H{\bf{\bar a}}_g^{( {\rm{full}} )}} |^2}$ for the fully-connected architecture and ${| {{\bf{h}}_{\Gamma ( g )}^H{\bf{\bar a}}_g^{( {\rm{sub}} )}} |^2}$ for the sub-connected architecture, separately, where $g = 1,2, \cdots ,G$. As a result, the $i$th element, where $i = 1,2, \cdots, N$, of the fully-connected analog precoding vector ${\bf{\bar a}}_g^{\left( {\rm{full}} \right)}$ can be expressed as
\begin{equation}\label{eq16}
{\bf{\bar a}}_g^{\left( {\rm{full}} \right)}\left( i \right) = \frac{1}{{\sqrt N }}{e^{j\frac{{2\pi \hat n}}{{{2^B}}}}},
\end{equation}
where
\begin{equation}\label{eq17}
\hat n = \mathop {\arg \min }\limits_{n \in \left\{ {0,1, \cdots, {2^B} - 1} \right\}} \left| {\rm{angle}}\left( {{\bf{h}}_{\Gamma \left( g \right)}\left( i \right)} \right) - \frac{{2\pi n}}{{{2^B}}} \right|.
\end{equation}
Similarly, the $i$th element of the sub-connected analog precoding vector ${\bf{\bar a}}_g^{\left( {\rm{sub}} \right)}$, where $i = ( {g - 1} )M + 1,( {g - 1} )M + 2, \cdots ,gM$, is
\begin{equation}\label{eq18}
{\bf{\bar a}}_g^{\left( {\rm{sub}} \right)}\left( i \right) = \frac{1}{{\sqrt M }}{e^{j\frac{{2\pi \hat n}}{{{2^B}}}}},
\end{equation}
where $\hat n$ is the same as that in (\ref{eq17}).

\subsection{User grouping}

After obtaining the analog precoding, the equivalent channel vectors for all $K$ users can be written as
\begin{equation}\label{eq19}
{\bf{\bar h}}_{k}^H = {\bf{h}}_{k}^H{\bf{A}},
\end{equation}
where $k = 1,2, \cdots, K$. Then, user grouping can be realized according to the correlation of equivalent channels. Specifically, user $m$ ($m \notin \Gamma $) can be classed as the ${\hat g}$th beam, where
\begin{equation}\label{eq20}
\hat g = \mathop {\arg \max }\limits_{g \in \left\{ {1,2, \cdots ,G} \right\}} \frac{{\left| {{\bf{\bar h}}_m^H{{{\bf{\bar h}}}_{\Gamma \left( g \right)}}} \right|}}{{{{\left\| {{{{\bf{\bar h}}}_m}} \right\|}_2}{{\left\| {{{{\bf{\bar h}}}_{\Gamma \left( g \right)}}} \right\|}_2}}}.
\end{equation}
In this way, users in the same beam will enjoy high correlation of equivalent channels, while the equivalent channels of users in different beams have low correlation owing to the proposed CHS algorithm, which is conducive to the inter-beam interference cancellation and thus the improvement of multiplexing gains.

\subsection{Digital precoding}

After analog precoding and user grouping, the equivalent channel vector for the $m$th user in the $g$th beam can be denoted as ${\bf{\bar h}}_{g,m}^H$ as introduced in Section II. Then, the design of digital precoding actually becomes a conventional MIMO-NOMA precoding problem to eliminate inter-beam interference. Without loss of generality, the low-complexity zero-forcing (ZF) precoding is adopted for digital precoding, according to the equivalent channel vectors of users having the highest equivalent channel gain in each beam~\cite{HPNOMA1}~\cite{HPNOMA2}.

Specifically, assuming that the $m_g$th user has the highest equivalent channel gain in the $g$th beam, we have
\begin{equation}\label{eq21}
{\bf{\bar H}} = \left[ {{{{\bf{\bar h}}}_{{m_1}}},{{{\bf{\bar h}}}_{{m_2}}}, \cdots ,{{{\bf{\bar h}}}_{{m_G}}}} \right].
\end{equation}
Then, the digital precoding matrix of size ${N_{\rm RF}} \times {N_{\rm RF}}$ can be generated by
\begin{equation}\label{eq22}
{\bf{\bar D}} = \left[ {{{{\bf{\bar d}}}_1},{{{\bf{\bar d}}}_2}, \cdots ,{{{\bf{\bar d}}}_G}} \right] = {\bf{\bar H}}{\left( {{{{\bf{\bar H}}}^H}{\bf{\bar H}}} \right)^{ - 1}}
\end{equation}
After normalizing, the digital precoding vector for the $g$th beam can be written as
\begin{equation}\label{eq23}
{{\bf{d}}_g} = \frac{{{{{\bf{\bar d}}}_g}}}{{{{\left\| {{\bf{A}}{{{\bf{\bar d}}}_g}} \right\|}_2}}}.
\end{equation}

Afterwards, the users in each beam will be reordered such that ${\| {{\bf{\bar h}}_{g,1}^H{{\bf{d}}_g}} \|_2} \ge {\| {{\bf{\bar h}}_{g,2}^H{{\bf{d}}_g}} \|_2} \ge  \cdots  \ge {\| {{\bf{\bar h}}_{g,| {{S_g}} |}^H{{\bf{d}}_g}} \|_2}$ for $g = 1,2, \cdots ,G$, which is assumed in Section II for SIC.

Up to now, user grouping and hybrid precoding have been carefully designed to obtain antenna array gains and multiplexing gains. In the next Section, the joint optimization of power allocation and power splitting will be investigated to maximize the achievable sum rate in (\ref{eq13}).

\section{Joint Optimization of Power Allocation and Power Splitting}\label{S4}

Although power allocation has been studied in existing MIMO-NOMA systems~\cite{NOMA3,NOMAPA1,NOMAPA2,NOMAPA3}, the joint optimization of power allocation and power splitting has not been considered. The introduction of power splitting factors will result in additional challenges for the joint optimization in mmWave massive MIMO-NOMA systems with SWIPT, since there exists not only the coupling of power allocation factors from different users, but also the coupling of power allocation and power splitting factors. On the other hand, the existing optimization methods for solving the joint optimization problem of power allocation and power splitting in MIMO systems with SWIPT cannot be directly used in MIMO-NOMA systems with SWIPT, where there are multiple groups and multiple users in each group, since both inter-group and intra-group interferences exist. Therefore, it is very difficult to obtain the optimal solutions. To solve this intractable problem, an iterative optimization algorithm is developed in this Section to obtain the sub-optimal solutions. Specifically, the joint power allocation and power splitting optimization problem can be formulated as
\begin{equation}\label{eq24}
\begin{array}{l}
\begin{aligned}
\mathop {\max }\limits_{\left\{ {{p_{g,m}}} \right\},\left\{ {{\beta _{g,m}}} \right\}} & \sum\limits_{g = 1}^G {\sum\limits_{m = 1}^{\left| {{S_g}} \right|} {{R_{g,m}}} }  \\
{\rm s.t.} \ \ \ {C_1}: & \ {p_{g,m}} \ge 0, \ \ \forall g,m, \\
{C_2}: & \ \sum\limits_{g = 1}^{G} {\sum\limits_{m = 1}^{\left| {{S_g}} \right|} {{p_{g,m}}} }  \le P_t,\\
{C_3}: & \ {R_{g,m}} \ge {R_{g,m}^{\min }}, \ \ \forall g,m, \\
{C_4}: & \ P_{g,m}^{\rm{EH}} \ge {P_{g,m}^{\min }}, \ \ \forall g,m,
\end{aligned}
\end{array}
\end{equation}
where ${{R_{g,m}}}$ is the achievable rate of the $m$the user in the $g$th beam as defined in (\ref{eq12}), the constraint $C_1$ indicates that the power allocated to each user must be positive, $C_2$ is the transmitted power constraint with $P_t$ being the maximum total transmitted power by the BS, $C_3$ is the data rate constraint for each user with ${R_{g,m}^{\min }}$ being the minimum data rate for the $m$the user in the $g$th beam, and $C_4$ is the EH QoS constraint for each user with ${P_{g,m}^{\min }}$ being the minimum harvested energy for the $m$the user in the $g$th beam. Note that the optimization problem (\ref{eq24}) is non-convex due to the non-convexity of the objective function, and the constraints $C_3$ as well as $C_4$.

To solve the non-convex problem (\ref{eq24}), an iterative optimization algorithm is developed. Particularly, according to the extension of the Sherman-Morrison-Woodbury formula~\cite{PA1}, i.e.,
\begin{equation}\label{eq25}
{\left( {{\bf{A}} + {\bf{BCD}}} \right)^{ - 1}} = {{\bf{A}}^{ - 1}} - {{\bf{A}}^{ - 1}}{\bf{B}}{\left( {{\bf{I}} + {\bf{CD}}{{\bf{A}}^{ - 1}}{\bf{B}}} \right)^{ - 1}}{\bf{CD}}{{\bf{A}}^{ - 1}},
\end{equation}
we have
\begin{equation}\label{eq26}
\begin{array}{l}
\begin{aligned}
&{\left( {1 + {\gamma _{g,m}}} \right)^{ - 1}} \\
 =& 1 - {p_{g,m}}\left\| {{\bf{\bar h}}_{g,m}^H{{\bf{d}}_g}} \right\|_2^2{\left( {{p_{g,m}}\left\| {{\bf{\bar h}}_{g,m}^H{{\bf{d}}_g}} \right\|_2^2 + {\xi _{g,m}}} \right)^{ - 1}},
\end{aligned}
\end{array}
\end{equation}
where $g = 1,2, \cdots ,G$ and $m = 1,2, \cdots ,\left| {{S_g}} \right|$.

On the other hand, let
\begin{equation}
\label{eq27}
\begin{array}{l}
\begin{aligned}
{{\tilde y}_{g,m}} =& {{\bf{\bar h}}_{g,m}^H{{\bf{d}}_g}\sqrt {{p_{g,m}}} {s_{g,m}} + {\bf{\bar h}}_{g,m}^H{{\bf{d}}_g}\sum\limits_{j = 1}^{m - 1} {\sqrt {{p_{g,j}}} {s_{g,j}}}} \\
& + {\bf{\bar h}}_{g,m}^H\sum\limits_{i \ne g} {\sum\limits_{j = 1}^{\left| {{S_i}} \right|} {{{\bf{d}}_i}\sqrt {{p_{i,j}}} {s_{i,j}}} }  + {v_{g,m}} + \frac{1}{{\sqrt {{\beta _{g,m}}} }}{u_{g,m}}.
\end{aligned}
\end{array}
\end{equation}
If the minimum mean square error (MMSE) detection is used to solve ${s_{g,m}}$ from ${{\tilde y}_{g,m}}$ in (\ref{eq27}), this detection problem can be formulated as
\begin{equation}\label{eq28}
c_{g,m}^o = \arg \mathop {\min }\limits_{{c_{g,m}}} {e_{g,m}},
\end{equation}
where
\begin{equation}\label{eq29}
{e_{g,m}} = {\rm{E}}\left\{ {{{\left| {{s_{g,m}} - {c_{g,m}}{{\tilde y}_{g,m}}} \right|}^2}} \right\}
\end{equation}
is the mean square error (MSE), and ${c_{g,m}}$ is the channel equalization coefficient. Substituting (\ref{eq27}) into (\ref{eq29}), we have
\begin{equation}\label{eq30}
\begin{array}{l}
\begin{aligned}
{e_{g,m}} =& 1 - 2{\rm{Re}}\left( {{c_{g,m}}\sqrt {{p_{g,m}}} {\bf{\bar h}}_{g,m}^H{{\bf{d}}_g}} \right) \\
 & + {\left| {{c_{g,m}}} \right|^2}\left( {{p_{g,m}}\left\| {{\bf{\bar h}}_{g,m}^H{{\bf{d}}_g}} \right\|_2^2 + {\xi _{g,m}}} \right).
\end{aligned}
\end{array}
\end{equation}
Then, by solving the partial derivatives of (\ref{eq28}) based on (\ref{eq30}), the optimal equalization coefficient $c_{g,m}^o$ can be obtained by
\begin{equation}\label{eq31}
c_{g,m}^o = {\left( {\sqrt {{p_{g,m}}} {\bf{\bar h}}_{g,m}^H{{\bf{d}}_g}} \right)^*}{\left( {{p_{g,m}}\left\| {{\bf{\bar h}}_{g,m}^H{{\bf{d}}_g}} \right\|_2^2 + {\xi _{g,m}}} \right)^{ - 1}}.
\end{equation}
Substituting (\ref{eq31}) into (\ref{eq30}), the MMSE can be written as
\begin{equation}\label{eq32}
e_{g,m}^o = 1 - {p_{g,m}}\left\| {{\bf{\bar h}}_{g,m}^H{{\bf{d}}_g}} \right\|_2^2{\left( {{p_{g,m}}\left\| {{\bf{\bar h}}_{g,m}^H{{\bf{d}}_g}} \right\|_2^2 + {\xi _{g,m}}} \right)^{ - 1}},
\end{equation}
which is equal to ${\left( {1 + {\gamma _{g,m}}} \right)^{ - 1}}$ in (\ref{eq26}), i.e., we have
\begin{equation}\label{eq33}
{\left( {1 + {\gamma _{g,m}}} \right)^{ - 1}} = \mathop {\min }\limits_{{c_{g,m}}} {e_{g,m}}.
\end{equation}
Then, the achievable rate of the $m$th user in the $g$th beam can be rewritten as
\begin{equation}\label{eq34}
{R_{g,m}} = {\log _2}\left( {1 + {\gamma _{g,m}}} \right) = \mathop {\max }\limits_{{c_{g,m}}} \left( { - {{\log }_2}{e_{g,m}}} \right).
\end{equation}
Note that in (\ref{eq34}), the polynomial division has been removed by using ${e_{g,m}}$ rather than ${\gamma _{g,m}}$, which significantly simplifies the objective function. Furthermore, to remove the log function in (\ref{eq34}), Proposition 1 is introduced~\cite{HPNOMA1}~\cite{NOMAPA1}.

\emph{Proposition 1}: Let $f\left( a \right) =  - \frac{{ab}}{{\ln 2}} + {\log _2} a  + \frac{1}{{\ln 2}}$ and $a$ be a positive real number, we have
\begin{equation}\label{eq35}
\mathop {\max }\limits_{a > 0} f\left( a \right) =  - {\log _2}b,
\end{equation}
where the optimal value of $a$ is ${a^o} = \frac{1}{b}$.

According to Proposition 1, we can rewrite (\ref{eq34}) as
\begin{equation}\label{eq36}
{R_{g,m}} = \mathop {\max }\limits_{{c_{g,m}}} \mathop {\max }\limits_{{a_{g,m}} > 0} \left( { - \frac{{{a_{g,m}}{e_{g,m}}}}{{\ln 2}} + {{\log }_2}{a_{g,m}} + \frac{1}{{\ln 2}}} \right).
\end{equation}
As a result, the optimization problem (\ref{eq24}) can be reformulated as
\begin{equation}\label{eq37}
\begin{array}{l}
\begin{aligned}
\mathop {\max }\limits_{\left\{ {{p_{g,m}}} \right\},\left\{ {{\beta _{g,m}}} \right\}} & \sum\limits_{g = 1}^G {\sum\limits_{m = 1}^{\left| {{S_g}} \right|} {\mathop {\max }\limits_{{c_{g,m}}} } \mathop {\max }\limits_{{a_{g,m}} > 0} \left( { - \frac{{{a_{g,m}}{e_{g,m}}}}{{\ln 2}} + {{\log }_2}{a_{g,m}}} \right)}    \\
& {\rm s.t.} \ \ \ {C_1}, {C_2}, {C_3}, {C_4}.
\end{aligned}
\end{array}
\end{equation}

To solve (\ref{eq37}), the iterative optimization algorithm is introduced to optimize $\{ {{c_{g,m}}} \}$, $\{ {{a_{g,m}}} \}$, and $\{ {{p_{g,m}}} \}$ as well as $\{ {\beta _{g,m}} \}$, separately. Specifically, given the optimal power allocation solution $\{ {p_{g,m}^{( {t - 1} )}} \}$ and the power splitting solution $\{ {\beta_{g,m}^{( {t - 1} )}} \}$ in the $(t-1)$th iteration, the optimal solution of $\{ {c_{g,m}^{( {t} )}} \}$ in the $t$th iteration can be obtained according to (\ref{eq31}), i.e.,
\begin{equation}\label{eq38}
\begin{array}{l}
\begin{aligned}
&c_{g,m}^{(t)}\\
=& {\left( {\sqrt {p_{g,m}^{\left( {t - 1} \right)}} {\bf{\bar h}}_{g,m}^H{{\bf{d}}_g}} \right)^*}{\left( {p_{g,m}^{\left( {t - 1} \right)}\left\| {{\bf{\bar h}}_{g,m}^H{{\bf{d}}_g}} \right\|_2^2 + \xi _{g,m}^{\left( {t - 1} \right)}} \right)^{ - 1}},
\end{aligned}
\end{array}
\end{equation}
where
\begin{equation}\label{eq39}
\begin{array}{l}
\begin{aligned}
{\xi _{g,m}}^{(t-1)} =& \left\| {{\bf{\bar h}}_{g,m}^H{{\bf{d}}_g}} \right\|_2^2\sum\limits_{j = 1}^{m - 1} {{p_{g,j}^{(t-1)}}}  \\
&+ \sum\limits_{i \ne g} {\left\| {{\bf{\bar h}}_{g,m}^H{{\bf{d}}_i}} \right\|_2^2} \sum\limits_{j = 1}^{\left| {{S_i}} \right|} {{p_{i,j}^{(t-1)}}} \\
& + \sigma _v^2 + \frac{{\sigma _u^2}}{{{\beta _{g,m}^{(t-1)}}}}.
\end{aligned}
\end{array}
\end{equation}
In the meanwhile, the optimal solution of $\{ {a_{g,m}^{( {t} )}} \}$ in the $t$th iteration can be calculated by
\begin{equation}\label{eq40}
a_{g,m}^{\left( t \right)} = \frac{1}{{e_{g,m}^{o\left( t \right)}}},
\end{equation}
where
\begin{equation}\label{eq41}
\begin{array}{l}
\begin{aligned}
&e_{g,m}^{o \left( t \right)}\\
 =& 1 - p_{g,m}^{\left( {t - 1} \right)}\left\| {{\bf{\bar h}}_{g,m}^H{{\bf{d}}_g}} \right\|_2^2{\left( {p_{g,m}^{\left( {t - 1} \right)}\left\| {{\bf{\bar h}}_{g,m}^H{{\bf{d}}_g}} \right\|_2^2 + \xi _{g,m}^{\left( {t - 1} \right)}} \right)^{ - 1}}.
\end{aligned}
\end{array}
\end{equation}
Then, the optimization problem (\ref{eq37}) can be simplified as
\begin{equation}\label{eq42}
\begin{array}{l}
\begin{aligned}
\mathop {\min }\limits_{\left\{ {p_{g,m}^{\left( t \right)}} \right\},\left\{ {\beta _{g,m}^{\left( t \right)}} \right\}} & \sum\limits_{g = 1}^G {\sum\limits_{m = 1}^{\left| {{S_g}} \right|} {a_{g,m}^{\left( t \right)}e_{g,m}^{\left( t \right)}} }     \\
{\rm s.t.} \ \ \ {C_1^{\left( t \right)}}: & \ {p_{g,m}^{\left( t \right)}} \ge 0, \ \ \forall g,m, \\
{C_2^{\left( t \right)}}: & \ \sum\limits_{g = 1}^{G} {\sum\limits_{m = 1}^{\left| {{S_g}} \right|} {{p_{g,m}^{\left( t \right)}}} }  \le P_t,\\
{C_3^{\left( t \right)}}: & \ {R_{g,m}^{\left( t \right)}} \ge {R_{g,m}^{\min }}, \ \ \forall g,m, \\
{C_4^{\left( t \right)}}: & \ P_{g,m}^{{\rm{EH}}{\left( t \right)}} \ge {P_{g,m}^{\min }}, \ \ \forall g,m,
\end{aligned}
\end{array}
\end{equation}
where
\begin{equation}\label{eq43}
\begin{array}{l}
\begin{aligned}
{e_{g,m}^{\left( t \right)}} =& 1 - 2{\rm{Re}}\left( {{c_{g,m}^{\left( t \right)}}\sqrt {{p_{g,m}^{\left( t \right)}}} {\bf{\bar h}}_{g,m}^H{{\bf{d}}_g}} \right) \\
 & + {\left| {{c_{g,m}^{\left( t \right)}}} \right|^2}\left( {{p_{g,m}^{\left( t \right)}}\left\| {{\bf{\bar h}}_{g,m}^H{{\bf{d}}_g}} \right\|_2^2 + {\xi _{g,m}^{\left( t \right)}}} \right).
\end{aligned}
\end{array}
\end{equation}
Furthermore, we introduce the variables $\{\tau _{g,m}^{( t )}\}$ such that $\tau _{g,m}^{( t )} \ge \frac{1}{{\beta _{g,m}^{( t )}}}$, and rewrite the objective function in (\ref{eq42}) as
\begin{equation}\label{eq44}
\mathop {\min }\limits_{\left\{ {p_{g,m}^{\left( t \right)}} \right\},\left\{ {\beta _{g,m}^{\left( t \right)}} \right\}} {\rm{ }}\sum\limits_{g = 1}^G {\sum\limits_{m = 1}^{\left| {{S_g}} \right|} {a_{g,m}^{\left( t \right)}\tilde e_{g,m}^{\left( t \right)}} } ,
\end{equation}
where
\begin{equation}\label{eq45}
\begin{array}{l}
\begin{aligned}
\tilde e_{g,m}^{\left( t \right)} = & 1 - 2{\rm{Re}}\left( {c_{g,m}^{\left( t \right)}{\bf{\bar h}}_{g,m}^H{{\bf{d}}_g}} \right)\sqrt {p_{g,m}^{\left( t \right)}} \\
& + {\left| {c_{g,m}^{\left( t \right)}} \right|^2}\left( {\left\| {{\bf{\bar h}}_{g,m}^H{{\bf{d}}_g}} \right\|_2^2\sum\limits_{j = 1}^m {p_{g,j}^{(t)}} }\right. \\
& \left. {+ \sum\limits_{i \ne g} {\left\| {{\bf{\bar h}}_{g,m}^H{{\bf{d}}_i}} \right\|_2^2} \sum\limits_{j = 1}^{\left| {{S_i}} \right|} {p_{i,j}^{(t)}}  + \sigma _u^2\tau _{g,m}^{\left( t \right)} + \sigma _v^2} \right),
\end{aligned}
\end{array}
\end{equation}
with the additional constraint
\begin{equation}\label{eq46}
{C_5^{\left( t \right)}}: \ \tau _{g,m}^{\left( t \right)} \ge \frac{1}{{\beta _{g,m}^{\left( t \right)}}}, \ \ \forall g,m.
\end{equation}

Intuitively, the objective function (\ref{eq44}) becomes convex for the optimization variables $\{ {p_{g,m}^{( {t} )}} \}$ and $\{ {\tau _{g,m}^{( {t} )}} \}$. At the same time, the constraint $C_3^{( t )}$ in (\ref{eq42}) can be also transformed into a convex constraint as
\begin{equation}\label{eq47}
\begin{array}{l}
\begin{aligned}
{\tilde C_3^{\left( t \right)}}: \ \ \ \ \ &\left\| {{\bf{\bar h}}_{g,m}^H{{\bf{d}}_g}} \right\|_2^2p_{g,m}^{\left( t \right)} - {\omega _{g,m}}\left\| {{\bf{\bar h}}_{g,m}^H{{\bf{d}}_g}} \right\|_2^2\sum\limits_{j = 1}^{m - 1} {p_{g,j}^{(t)}}  \\
& - {\omega _{g,m}}\sum\limits_{i \ne g} {\left\| {{\bf{\bar h}}_{g,m}^H{{\bf{d}}_i}} \right\|_2^2} \sum\limits_{j = 1}^{\left| {{S_i}} \right|} {p_{i,j}^{(t)}}  - {\omega _{g,m}}\sigma _u^2\tau _{g,m}^{\left( t \right)} \\
\ge & {\omega _{g,m}}\sigma _v^2,
\end{aligned}
\end{array}
\end{equation}
where ${\omega _{g,m}} = {2^{R_{g,m}^{\min }}} - 1$. Nevertheless, the constraints ${C_4^{\left( t \right)}}$ and ${C_5^{\left( t \right)}}$ in (\ref{eq42}) are still non-convex due to the multi-variable coupling. To make it solvable, another variables $\{\mu _{g,m}^{( t )}\}$ are introduced such that
\begin{equation}\label{eq48}
{C_6^{\left( t \right)}}: \ \mu _{g,m}^{\left( t \right)} \ge \frac{{P_{g,m}^{\min }}}{{\eta \left( {1 - \beta _{g,m}^{\left( t \right)}} \right)}}, \ \ \forall g,m.
\end{equation}
Then, the constraint ${C_4^{\left( t \right)}}$ in (\ref{eq42}) can be rewritten as
\begin{equation}\label{eq49}
{\tilde C_4^{\left( t \right)}}: \ \sum\limits_{i = 1}^G {\sum\limits_{j = 1}^{\left| {{S_i}} \right|} {\left\| {{\bf{\bar h}}_{g,m}^H{{\bf{d}}_i}} \right\|_2^2p_{i,j}^{\left( t \right)}} }  + \sigma _v^2 \ge \mu _{g,m}^{\left( t \right)}, \ \ \forall g,m,
\end{equation}
which becomes a convex constraint.

To deal with the non-convex constraints ${C_5^{\left( t \right)}}$ in (\ref{eq46}) and ${C_6^{\left( t \right)}}$ in (\ref{eq48}), we transform them into matrix form according to the Schur complement lemma~\cite{SWIPT8}, i.e.,
\begin{equation}\label{eq50}
{\tilde C_5^{\left( t \right)}}: \ \left[ {\begin{array}{*{20}{c}}
{\tau _{g,m}^{\left( t \right)}}&1\\
1&{\beta _{g,m}^{\left( t \right)}}
\end{array}} \right] \ge {\bf{0}}, \ \ \forall g,m,
\end{equation}
and
\begin{equation}\label{eq51}
{\tilde C_6^{\left( t \right)}}: \ \left[ {\begin{array}{*{20}{c}}
{\mu _{g,m}^{\left( t \right)}}&{\sqrt {{{P_{g,m}^{\min }} \mathord{\left/
 {\vphantom {{P_{g,m}^{\min }} \eta }} \right.
 \kern-\nulldelimiterspace} \eta }} }\\
{\sqrt {{{P_{g,m}^{\min }} \mathord{\left/
 {\vphantom {{P_{g,m}^{\min }} \eta }} \right.
 \kern-\nulldelimiterspace} \eta }} }&{1 - \beta _{g,m}^{\left( t \right)}}
\end{array}} \right] \ge {\bf{0}}, \ \ \forall g,m.
\end{equation}

As a result, the optimization problem (\ref{eq42}) can be reformulated as
\begin{equation}\label{eq52}
\begin{array}{l}
\begin{aligned}
\mathop {\min }\limits_{\left\{ {p_{g,m}^{\left( t \right)}} \right\},\left\{ {\beta _{g,m}^{\left( t \right)}} \right\}} & \sum\limits_{g = 1}^G {\sum\limits_{m = 1}^{\left| {{S_g}} \right|} {a_{g,m}^{\left( t \right)}\tilde e_{g,m}^{\left( t \right)}} }    \\
{\rm s.t.} \ \ \ & {C_1^{\left( t \right)}}, {C_2^{\left( t \right)}}, {\tilde C_3^{\left( t \right)}}, {\tilde C_4^{\left( t \right)}}, {\tilde C_5^{\left( t \right)}}, {\tilde C_6^{\left( t \right)}},
\end{aligned}
\end{array}
\end{equation}
which is a standard convex optimization problem, and can be solved by numerical convex program solvers~\cite{cvx}.

By iteratively solving the optimal values of $\{ {{c_{g,m}}} \}$, $\{ {{a_{g,m}}} \}$, and $\{ {{p_{g,m}}} \}$ as well as $\{ {\beta _{g,m}} \}$ via (\ref{eq38}), (\ref{eq40}), and (\ref{eq52}), separately, we can obtain the final power allocation solution $\{ {{p_{g,m}^o}} \}$ and power splitting solution $\{ {\beta _{g,m}^o} \}$ with the maximum iteration times $T_{\rm{max}}$. Particularly, since the obtained $\{ c_{m,n}^{( t )} \}$, $\{ a_{m,n}^{( t )} \}$, and $\{ p_{m,n}^{( t )} \}$ as well as $\{ \beta _{m,n}^{( t )} \}$ are optimal solutions in the $t$th iteration, iteratively updating these variables will increase or maintain the value of the objective function in (\ref{eq37})~\cite{NOMAPA1}. As a result, the proposed iterative optimization algorithm for joint power allocation and power splitting will converge to at least a local optimal solution.

In the meanwhile, the proposed joint optimization algorithm enjoys a polynomial complexity. Specifically, in each iteration, the complexity for obtaining the optimal $\{ c_{m,n} \}$ in (\ref{eq38}) and $\{ a_{m,n} \}$ in (\ref{eq40}) is linear to the number of users, i.e., $\mathcal {O} (K)$. The convex optimization problem (\ref{eq52}) can be solved with a worst-case
complexity of ${\cal O}({T_{\max }}{K^{4.5}}{\log _2}({1 \mathord{\left/ {\vphantom {1 \varepsilon }} \right. \kern-\nulldelimiterspace} \varepsilon }))$ given a solution accuracy $\varepsilon > 0 $~\cite{cvx2}. Therefore, the computational complexity of the proposed iterative optimization algorithm is at most ${\cal O}({T_{\max }}{K^{4.5}}{\log _2}({1 \mathord{\left/ {\vphantom {1 \varepsilon }} \right. \kern-\nulldelimiterspace} \varepsilon }))$.

%
%
%
%
%
%
%
%
%
%

\section{Simulation Results}\label{S5}

The performance in terms of spectrum efficiency and energy efficiency of the mmWave massive MIMO-NOMA systems with SWIPT, including both the fully-connected HP and the sub-connected HP, proposed in this paper is evaluated via simulations. Specifically, the simulation parameters are described as follows: The system bandwidth is assumed to be 1 Hz, which coincides to the achievable rate in (12). The BS is equipped with an ULA of $N=64$ antennas and $N_{\rm{RF}}=4$ RF chains to simultaneously serve $K \ge N_{\rm{RF}}$ users. All the $K$ uses are grouped into $G=N_{\rm{RF}}=4$ beams, and there are more than one user in each beam. For the $m$th user in the $g$th beam, the channel vector is generated based on (\ref{eq4}), where we assume: 1) $L_{g,m}=3$, including one line-of-sight (LoS) component and two non-line-of-sight (NLoS) components; 2) ${\alpha _{g,m}^{\left( 1 \right)}} \sim \mathcal{CN}(0,1)$, and ${\alpha _{g,m}^{\left( l \right)}} \sim \mathcal{CN}(0,10^{ - 1})$ for $2 \le l \le L_{g,m}$; 3) ${\varphi _{g,m}^{\left( l \right)}}$ and ${\theta _{g,m}^{\left( l \right)}}$ follow the uniform distribution $\mathcal{U}(- \pi ,\pi)$ for $1 \le l \le L_{g,m}$. $B=4$ bits quantized phase shifters are adopted, and the signal-to-noise ratio (SNR) is defined as ${{{P_t}} \mathord{\left/ {\vphantom {{{P_t}} {\sigma _v^2}}} \right. \kern-\nulldelimiterspace} {\sigma _v^2}}$~\cite{HPMIMO3}. The maximum transmitted power $P_t=30$ mW, the minimal achievable rate for each user is ${R_{fm}}/10$, where ${R_{fm}}$ is the minimal achievable rate among all users by using fully digital ZF precoding, and the minimal harvested energy for each user is 0.1 mW.

In this paper, the spectrum efficiency is defined as the achievable sum rate in (\ref{eq13}), and the energy efficiency is defined as the ratio between the achievable sum rate and the total power consumption~\cite{HPMIMO2}, i.e.,
\begin{equation}\label{eq53}
{\rm{EE}}  = \frac{{{R_{\rm sum}}}}{{P_{tr} + {N_{\rm RF}}{P_{\rm RF}}+{N_{\rm PS}}{P_{\rm PS}}+P_{\rm BB}}} \ ({\rm bps/Hz/W}),
\end{equation}
where ${P_{tr}} = \sum\limits_{g = 1}^G {\sum\limits_{m = 1}^{\left| {{S_g}} \right|} {{p_{g,m}}} } $ is the total transmitted power, $P_{\rm RF}$ is the power consumed by each RF chain, $P_{\rm PS}$ is the power consumption of each phase shifter, and $P_{\rm BB}$ is the baseband power consumption. Particularly, we adopt the typical values $P_{\rm RF} = 300$ mW, $P_{\rm PS} = 40$ mW (4-bit phase shifter), and $P_{\rm BB} = 200$ mW~\cite{HPMIMO3}. $N_{\rm PS}$ is the number of phase shifters, which is equal to $NN_{\rm{RF}}$ for the fully-connected HP and $N$ for the sub-connected HP.

In the simulations, we consider the following five typical mmWave massive MIMO systems with SWIPT for comparison: (1) ``SWIPT-Fully digital ZF Precoding'', where each antenna is connected to one RF chain, and ZF precoding is adopted; (2) ``SWIPT-Fully-Connected HP-NOMA'', where fully-connected HP architecture is used in the proposed mmWave massive MIMO-NOMA systems with SWIPT; (3) ``SWIPT-Fully-Connected HP-OMA'', where the system model is similar with ``SWIPT-Fully-Connected HP-NOMA'', while OMA is performed for users in each beam. Particularly, we represent OMA with FDMA, where users in the same beam are allocated with equal bandwidth; (4) ``SWIPT-Sub-Connected HP-NOMA'', where sub-connected HP architecture is used in the proposed mmWave massive MIMO-NOMA systems with SWIPT; (5) ``SWIPT-Sub-Connected HP-OMA'', where the system model is similar with ``SWIPT-Sub-Connected HP-NOMA'', while OMA is performed for users in each beam.

\begin{figure}[tp]
\begin{center}
\includegraphics[width=1\linewidth]{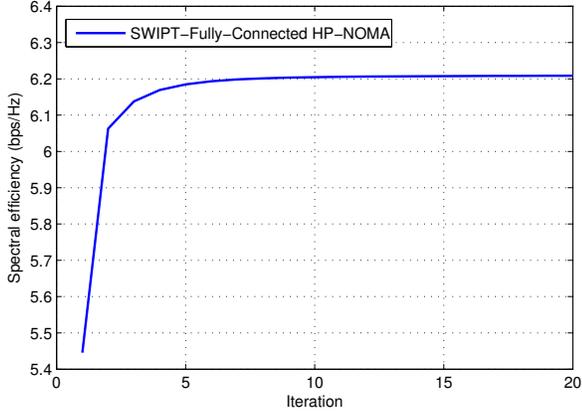} \caption{Spectrum efficiency of fully-connected architecture against the number of iterations for the joint power allocation and power splitting optimization.}
\end{center}
\end{figure}

\begin{figure}[tp]
\begin{center}
\includegraphics[width=1\linewidth]{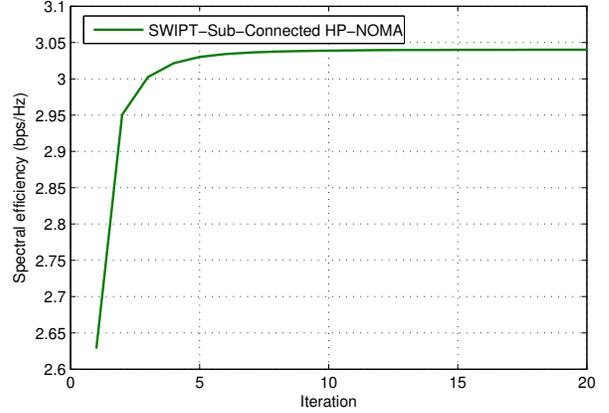} \caption{Spectrum efficiency of sub-connected architecture against the number of iterations for the joint power allocation and power splitting optimization.}
\end{center}
\end{figure}

Fig. 2 and Fig. 3 show the convergence of the proposed iterative algorithm for joint power allocation and power splitting optimization in Section IV for the fully-connected HP and sub-connected HP, separately, where the number of users is set as $K = 6$, and SNR = 0 dB. As shown in Fig. 2 and Fig. 3, the spectrum efficiency tends to be stable after 10 times of iteration, which verifies the convergence of the proposed iterative algorithm as discussed in Section IV. In the following simulations, the number of iteration times for the proposed iterative optimization algorithm is set as 10.

\begin{figure}[tp]
\begin{center}
\includegraphics[width=1\linewidth]{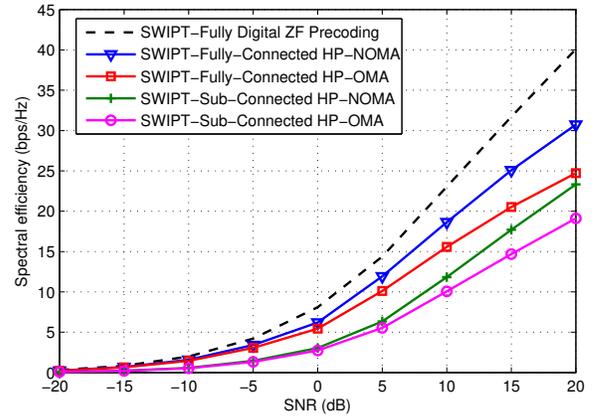} \caption{Spectrum efficiency against SNR.}
\end{center}
\end{figure}

Fig. 4 shows the spectrum efficiency against SNR of the considered five schemes, where the number of users is $K = 6$. We can find that the proposed mmWave massive MIMO-NOMA systems with SWIPT can achieve higher spectrum efficiency than that of mmWave massive MIMO-OMA systems with SWIPT, either for the fully-connected HP or the sub-connected HP, since NOMA can achieve higher spectrum efficiency than that of OMA~\cite{NOMA1}. It is intuitive that the fully digital MIMO can achieve the best spectrum efficiency as shown in Fig. 4, since $N$ RF chains are used to serve all users to fully exploit the multiplexing gains. On the other hand, the fully-connected HP can achieve higher spectrum efficiency than that of the sub-connected HP as discussed in Section II, since the full array gain can be exploited by every RF chain in the fully-connected HP.

\begin{figure}[tp]
\begin{center}
\includegraphics[width=1\linewidth]{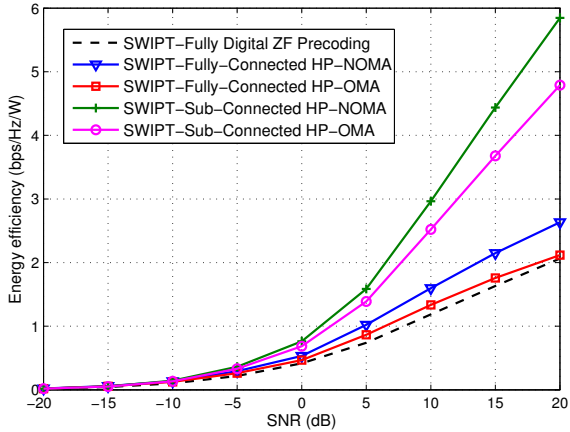} \caption{Energy efficiency against SNR.}
\end{center}
\end{figure}

Fig. 5 shows the energy efficiency against SNR, where the number of users is also $K = 6$. We can find that the proposed mmWave massive MIMO-NOMA systems with SWIPT can achieve higher energy efficiency than both mmWave massive MIMO-OMA systems with SWIPT and fully digital MIMO systems with SWIPT. Particularly, the number of RF chains is equal to the number of BS antennas in fully digital MIMO systems, which leads to very high energy consumption, e.g., 300 mW for each RF chain. On the contrary, the number of RF chains is much smaller than the number of antennas in the proposed mmWave massive MIMO-NOMA systems with SWIPT. Therefore, the energy consumption caused by the RF chains can be significantly reduced compared to the fully digital MIMO systems. In addition, we can see from Fig. 5 that the sub-connected HP can achieve higher energy efficiency than that of the fully-connected HP, since a less number of phase shifters is adopted in the sub-connected HP.

\begin{figure}[tp]
\begin{center}
\includegraphics[width=1\linewidth]{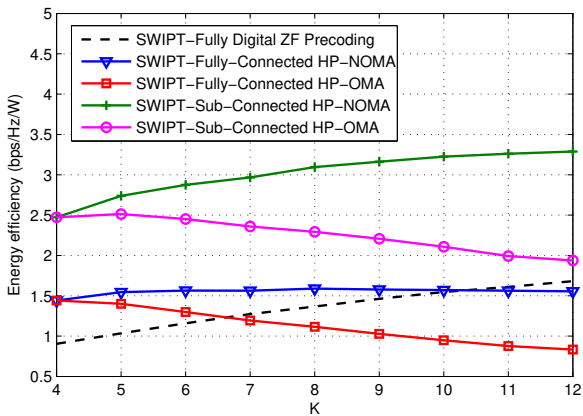} \caption{Energy efficiency against the number of users.}
\end{center}
\end{figure}

The performance comparison in terms of energy efficiency against the number of users is shown in Fig. 6, where SNR is set as 10 dB. We can see that when the sub-connected HP is adopted, the energy efficiency of the proposed mmWave massive MIMO-OMA systems with SWIPT is higher than all of other schemes even the number of users is very large.

\section{Conclusions}\label{S7}

In this paper, we propose to apply SWIPT in HP-based mmWave massive MIMO-NOMA systems to achieve a trade-off between spectrum efficiency and energy efficiency. To enable the spectrum- and energy-efficient systems, user grouping, hybrid precoding, power allocation, and power splitting are carefully designed. Specifically, the CHS algorithm is first proposed to select one user for each beam as the cluster head, and then the analog precoding is designed according to the selected cluster heads for all beams. After that, user grouping is performed based on the correlation of users' equivalent channels. Then, the digital precoding is designed by selecting users with the strongest equivalent channel gain in each beam. Furthermore, the joint optimization of power allocation and power splitting is proposed to maximize the achievable sum rate, and an iterative optimization algorithm is developed to solve the non-convex optimization problem. Simulation results show that the proposed mmWave massive MIMO-NOMA systems with SWIPT can achieve higher spectrum and energy efficiency compared with mmWave massive MIMO-OMA systems with SWIPT. In the future, we will consider more sophisticated hybrid precoding design for the proposed mmWave massive MIMO-NOMA systems with SWIPT to further improve the performance.

\begin{IEEEbiography}[{\includegraphics[width=1in,height=1.2in,keepaspectratio]{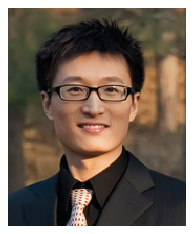}}]{Linglong Dai}
(M'11-SM'14) received the B.S. degree from Zhejiang University in 2003, the M.S. degree (with the highest honor) from the China Academy of Telecommunications Technology in 2006, and the Ph.D. degree (with the highest honor) from Tsinghua University, Beijing, China, in 2011. From 2011 to 2013, he was a Post-Doctoral Research Fellow with the Department of Electronic Engineering, Tsinghua University, where he was an Assistant Professor from 2013 to 2016 and has been an Associate Professor since 2016. He coauthored the book ``mmWave Massive MIMO: A Paradigm for 5G'' (Academic Press, Elsevier, 2016). He has published over 60 IEEE journal papers and over 40 IEEE conference papers. He also holds 16 granted patents. His current research interests include massive MIMO, millimeter-wave communications, NOMA, sparse signal processing, and machine learning for wireless communications. He has received five conference Best Paper Awards at the IEEE ICC 2013, the IEEE ICC 2014, the IEEE ICC 2017, the IEEE VTC 2017-Fall, and the IEEE ICC 2018. He has also received the Tsinghua University Outstanding Ph.D. Graduate Award in 2011, the Beijing Excellent Doctoral Dissertation Award in 2012, the China National Excellent Doctoral Dissertation Nomination Award in 2013, the URSI Young Scientist Award in 2014, the IEEE Transactions on Broadcasting Best Paper Award in 2015, the Second Prize of Science and Technology Award of China Institute of Communications in 2016, the Electronics Letters Best Paper Award in 2016, the IEEE Communications Letters Exemplary Editor Award in 2017, the National Natural Science Foundation of China for Outstanding Young Scholars in 2017, and the IEEE ComSoc Asia-Pacific Outstanding Young Researcher Award in 2017. He currently serves as an Editor of the IEEE Transactions on Communications, the IEEE Transactions on Vehicular Technology, and the IEEE Communications Letters.
\end{IEEEbiography}

\begin{IEEEbiography}[{\includegraphics[width=1in,height=1.2in,keepaspectratio]{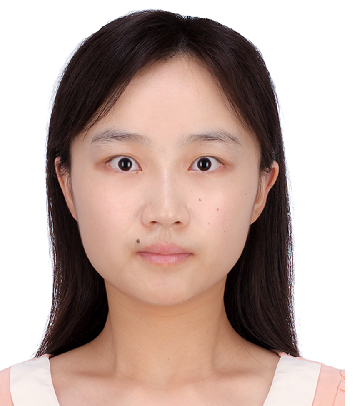}}]{Bichai Wang}
(S'15) received her B.S. degree in Electronic Engineering from Tsinghua University, Beijing, China, in 2015. She is currently working towards the Ph.D. degree in the Department of Electronic Engineering, Tsinghua University, Beijing, China. Her research interests are in wireless communications, with the emphasis on non-orthogonal multiple access, mmWave massive MIMO, and deep learning-based wireless communications. She has received the Freshman Scholarship of Tsinghua University in 2011, the Academic Merit Scholarships of Tsinghua University in 2012, 2013, and 2014, respectively, the Excellent Thesis Award of Tsinghua University in 2015, the National Scholarship in 2016, the IEEE VTC'17 Fall Best Student Paper Award in 2017, the IEEE Transactions on Communications Exemplary Reviewer Award in 2017, and the first prize of the 13th China Graduate Electronics Design Contest in 2018.
\end{IEEEbiography}

\begin{IEEEbiography}[{\includegraphics[width=1in,height=1.2in,keepaspectratio]{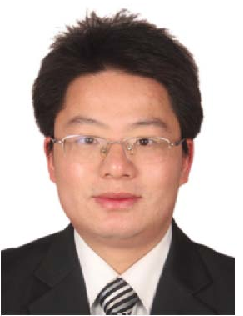}}]{Mugen Peng}
(M'05--SM'11) received the Ph.D. degree in communication and information systems from the Beijing University of Posts and Telecommunications (BUPT), Beijing, China, in 2005. Afterward, he joined BUPT, where he has been a Full Professor with the School of Information and Communication Engineering since 2012. He leads a Research Group focusing on wireless transmission and networking technologies with the Key Laboratory of Universal Wireless Communications (Ministry of Education), BUPT. His main research areas include wireless communication theory, radio signal processing, and convex optimizations, with a particular interests in cooperative communication, self-organization networking, heterogeneous networking, cloud communication, and internet of things. He has authored/coauthored over 100 refereed IEEE journal papers and over 200 conference proceeding papers.

Dr. Peng was a recipient of the 2018 Heinrich Hertz Prize Paper Award, the 2014 IEEE ComSoc AP Outstanding Young Researcher Award, and the best paper award in IEEE WCNC 2015, WASA 2015, GameNets 2014, IEEE CIT 2014, ICCTA 2011, IC-BNMT 2010, and IET CCWMC 2009. He received the First Grade Award of the Technological Invention Award in the Ministry of Education of China, and the First Grade Award of Technological Invention Award from the China Institute of Communications. He is on the Editorial/Associate Editorial Board of the IEEE Communications Magazine, IEEE Access, IEEE Internet of Things Journal, IET Communications, and China Communications.
\end{IEEEbiography}

\begin{IEEEbiography}[{\includegraphics[width=1in,height=1.2in,keepaspectratio]{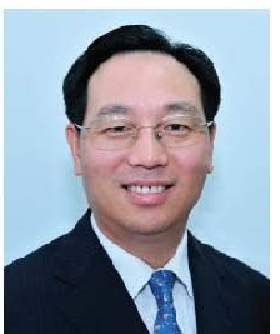}}]{Shanzhi Chen}
(SM'04) received the bachelor¡¯s degree from Xidian University in 1991 and the Ph.D. degree from the Beijing University of Posts and Telecommunications, China, in 1997. He joined the Datang Telecom Technology and Industry Group and China Academy of Telecommunication Technology (CATT) in 1994, and has served as EVP Research and Development since 2008. He is currently the Director of the State Key Laboratory of Wireless Mobile Communications, CATT, where he conducted research and standardization on 4G TD-LTE and 5G. He has authored and co-authored four books [among them the textbook Mobility Management: Principle, Technology and Applications, (Springer Press)], 17 book chapters, approximately 100 journal papers, 50 conference papers, and over 50 patents in these areas. He has contributed to the design, standardization, and development of 4G TD-LTE and 5G mobile communication systems. His current research interests include 5G mobile communications, network architectures, vehicular communication networks, and Internet of things. He served as a member and a TPC Chair and of many international conferences. His achievements have received multiple top awards by China central government and honors, especially, the Grand Prize of National Award for Scientific and Technological Progress, China, in 2016 (This Grand Prize is the highest category and in some years, it leaves with no winners due to its high standard). He is the Editor of the IEEE Network and the IEEE Internet of Things, and the Guest Editor for the IEEE Wireless Communications, the IEEE Communications Magazine, and the IEEE Transactions on Vehicular Technology.
\end{IEEEbiography}

\end{document}